\documentclass[namedreferences]{kluwer}    

\newdisplay{guess}{Conjecture}

\begin{document}
\begin{article}

\begin{opening}         


\title{The velocity field of young stars\\ in the solar neighbourhood}


\author{J. \surname{Torra}}
\author{D. \surname{Fern\'andez}}
\author{F. \surname{Figueras}}
\institute{Departament d'Astronomia i Meteorologia, Universitat de
           Barcelona, Av. Diagonal 647, E-08028 Barcelona, Spain}

\author{F. \surname{Comer\'on}}
\institute{European Southern Observatory, Karl-Schwarzschild-Strasse 2,
           D-85748 Garching bei M\"unchen, Germany}


\runningauthor{J. Torra et al.}
\runningtitle{The velocity field of young stars in the solar 
              neighbourhood}
\begin{ao}
\\J. Torra
\\e-mail: jordi@am.ub.es
\end{ao}



\begin{abstract}
A sample of O- and B-type stars with Hipparcos astrometric data, ages
computed from Str\"omgren photometry and radial velocities, has been used
to characterize the structure, age and kinematics of the Gould Belt
system. The local spiral structure of our galaxy is determined from this
sample, and also from a sample of Hipparcos Cepheid stars.

The Gould Belt, with an orientation with respect to the galactic plane of
$i_{\mathrm{G}} = 16$-$22^\circ$ and $\Omega_{\mathrm{G}} =
275$-$295^\circ$, extends up to a distance of 600 pc from the Sun. Roughly
the 60-65\% of the O and B stars younger than 60 Myr in the solar
neighbourhood belong to this structure. Our results indicate that the
kinematical behaviour of this system is complex, with an expansion motion
in the solar neighbourhood ($R <$ 300 pc).

In the frame of the Lin's theory, and analysing the O and B stars further
than 600 pc and the Cepheids, we found a galactic spiral structure
characterized by a 4-arm spiral pattern with the Sun located at
$\psi_\odot = 350$-$355 \pm 30^\circ$ -- near the Sagittarius-Carina arm
-- and outside the corotation circle. The angular rotation speed of the
spiral pattern was found to be $\Omega_{\mathrm{p}} = 31$-$32 \pm 4$ km
s$^{-1}$ kpc$^{-1}$. \end{abstract}


\end{opening}

\section{The working samples}

\subsection{Sample of O and B stars}

Our initial sample contains 6922 O and B stars. The astrometric data come
from Hipparcos Catalogue \cite{ESA97}, radial velocities from the
compilation of \inlinecite{Grenier} and Str\"omgren photometry from
\inlinecite{Hauck et al.} catalogue (see details in \opencite{Fernandez}).
Photometric distances were computed from \inlinecite{Crawford}
calibration. Close binaries, high amplitude variables and peculiar stars
were rejected. Once the trigonometric and photometric distances were
known, that with small relative error was chosen. Individual ages were
computed from the models of \inlinecite{Bressan et al.} following the
interpolation algorithm described in \inlinecite{Asiain et al.}.

The sample with distances, proper motions and ages contains 2468 stars,
whereas the subsample with available radial velocities contains 1789
stars.

\subsection{Sample of Cepheid stars}

The initial sample contains all the Hipparcos classical Cepheids, whereas
overtone Cepheids were eliminated. Astrometric data were taken from the
Hipparcos Catalogue \cite{ESA97} and radial velocities from the Hipparcos
Input Catalogue \cite{ESA92}. Using Hipparcos data, \inlinecite{Luri et
al.98} derived individual estimates of luminosities and distances for
these stars through the LM method \cite{Luri et al.96}.

The sample with known distances and proper motions contains 207 stars. The
subsample with known radial velocities contains 99 stars.

\section{Characterizing the Gould Belt}

\subsection{Structure parameters and age}

Our model assumes that both the Gould and galactic belts trace two great
circles in the celestial sphere. The decrease of star density with the
angular distance to each belt was assumed to follow a gaussian law, the
standard deviation being the angular halfwidth of the belt. The resolution
procedure was based on the maximum likelihood method. An iterative
procedure was implemented to minimize the dependence of the final results
on the initial values (see details in \opencite{Comeron}, and
\opencite{Torra et al.}).

The Gould Belt structure is clearly present in the subsamples of young
stars with $R \le$ 600 pc. The orientation parameters were found to be
$i_{\mathrm{G}} = 16$-$22^\circ$ and $\Omega_{\mathrm{G}} =
275$-$295^\circ$, depending on the age and the distance interval
considered. For stars with $\tau \le$ 60 Myr and $R \le$ 600 pc, the
fraction of them belonging to the Gould Belt was found to be 60-66\%. This
percentage decreased to 42-44\% when stars with 60 $< \tau \le$ 90 Myr
were considered. We estimate an age of the Gould Belt inside the interval
60-90 Myr. The angular halfwidth of the belts were found to be 6-8$^\circ$
and 23-26$^\circ$ for the Gould and galactic belts, respectively.

\subsection{The Gould Belt kinematics}

The Oort constants $A$, $B$, $C$ and $K$ were derived using the
first-order development of the galactic velocity field. A combined
solution was applied, solving simultaneously radial velocity and proper
motion equations, and rejecting high residual stars.

In the region with $R >$ 600 pc, the stellar kinematics is dominated by
the differential galactic rotation, since the Oort constants were found to
be $A = 13.0 \pm 0.5$ km s$^{-1}$ kpc$^{-1}$, $B = -12.2 \pm 0.5$ km
s$^{-1}$ kpc$^{-1}$, $C = 0.2 \pm 0.5$ km s$^{-1}$ kpc$^{-1}$ and $K =
-2.0 \pm 0.5$ km s$^{-1}$ kpc$^{-1}$. Contrary to that, in the region with
$R \le$ 600 pc, the Gould Belt defines the kinematics of the youngest
stars ($\tau \leq$ 60-90 Myr), producing a decrease of the $A$ and $B$
Oort constants ($A \approx$ 6-8 km s$^{-1}$ kpc$^{-1}$, $B \approx
-$(22-14) km s$^{-1}$ kpc$^{-1}$) and an increase of $C$ and $K$ ($C
\approx$ 6-9 km s$^{-1}$ kpc$^{-1}$, $K \approx$ 4-7 km s$^{-1}$
kpc$^{-1}$). This peculiar kinematics was also found when those stars
belonging to the Sco-Cen and Ori OB1 complexes were eliminated. Therefore,
these associations are not the only responsibles for these peculiarities.
Through the analysis of the variation of the Oort constants with the age,
a kinematic age of the Gould Belt inside the interval 60-90 Myr was
inferred, in good agreement with previous estimations from the spatial
distribution of stars.

The study of the residual velocity field of the youngest stars shows that
it cannot be explained as an expansion from a point \cite{Olano} or a line
\cite{Comeron et al.}. Moreover, the expansion motion classically
attributed to the Gould Belt seems to be due to the nearest stars ($R \le$
300 pc). In the region 300 $< R \le$ 600 pc, only Per OB2 has a clear
residual motion away from the Sun.

\section{Galactic spiral structure}

The spiral structure of the Galaxy was studied developing the galactic
velocity field projected on the galactic disk, taking into account the
contributions of the solar motion, differential galactic rotation (up to
second-order approximation) and spiral arm kinematics (modelled in the
frame of Lin's theory). See details of the model in \inlinecite{Fernandez
et al.}.

Coherent results for O-B stars ($R >$ 600 pc) and Cepheids ($R <$ 4000 pc)
are obtained when adopting a 4-arm spiral pattern with a pitch angle of
$-14^\circ$ \cite{Amaral et al.}, with the Sun placed at 7.1 kpc from the
galactic center and the circular velocity being equal to 184 km s$^{-1}$
\cite{Olling et al.}. Other values of these parameters have been also
discussed in \inlinecite{Fernandez et al.}. From a least squares fit we
derived that the Sun is located in an arm, between its center and the
outer edge ($\psi_\odot = 350$-$355 \pm 30^\circ$), near the
Sagittarius-Carina arm and outside the corotation circle, placed at
$\varpi_{\mathrm{cor}} = 5.9$-$6.1 \pm 0.7$ kpc. The angular rotation
velocity of the spiral structure was found to be $\Omega_{\mathrm{p}} =
31$-$32 \pm 4$ km s$^{-1}$ kpc$^{-1}$. Following \inlinecite{Lindblad}, we
checked that the obtained spiral structure pattern can justify the
difference found in the Oort constants (except in the case of $B$) between
solutions using nearby O-B stars not belonging to the Gould Belt ($R \le$
600 pc and $\tau >$ 90 Myr) and O-B stars at $R > 600$ pc (see
\opencite{Torra et al.}).

\acknowledgements

This work has been suported by the CICYT under contract ESP 97-1803 and by
the PICS programme (CIRIT). DF acknowledges the FRD grant of the
Universitat de Barcelona.

\end{article}
\end{document}